\documentstyle[aas2pp4]{article}

\lefthead{PAGE}
\righthead{Fast Cooling of Neutron Stars}

\newcommand{\be}{\begin{equation}} 
\newcommand{\ee}{\end{equation}} 
\newcommand{\Msol}{\mbox{$M_{\odot}\;$}}
\newcommand{\Msun}{\mbox{$M_{\odot}\;$}}
\newcommand{\tnm}{\tablenotemark}
\newcommand{\tnt}{\tablenotetext}
 
\begin{document}
 
\title{FAST COOLING OF NEUTRON STARS:
       SUPERFLUIDITY \\ vs. HEATING AND ACCRETED ENVELOPE}
\author{Dany Page}
\affil{Instituto de Astronom\'{\i}a, UNAM, 
       Apdo Postal 70-264,
       04510 M\'{e}xico D.F., M\'{e}xico. page@astroscu.unam.mx}

\begin{abstract}

It is generally considered that the neutron star cooling scenarios involving
fast neutrino emission, from a kaon or pion condensate, quark matter, or the
direct Urca process, require the presence of baryon pairing in the central core
of the star to control the strong neutrino emission and produce surface
temperatures compatible with observations.
I show here that within the kaon condensate scenario {\em pairing is not 
necessary} if: 
1) the equation of state is stiff enough for the star to have a thick
crust in which sufficient friction can occur to heat the star and 2) a
thin layer, of mass $\Delta M$ larger than $\sim 10^{-12}$ \Msol, of
light elements (H and He) is present at the stellar surface.
The effect of the light elements is to increase the heat flow and thus produce
a higher surface temperature.
Both the occurrence of heating and the presence of H and/or He at the surface
(deposited during the late post-supernova accretion) can possibly be confirmed
or infirmed by future observations.

\end{abstract}


\bigskip
\bigskip
Submitted to {\em The Astrophysical Journal Letters}

\newpage 
\section{INTRODUCTION}

There are indications that some young pulsars may have surface temperatures
lower than what is predicted by the `standard' model of neutron star cooling
(see, e.g.,  Nomoto \& Tsuruta 1986).
Alternative models with fast neutrino emission (see, e.g., Page 1994), however,
predict temperatures much lower than observed.
It had been proposed (Page \& Baron 1990; Page \& Applegate 1992) that baryon 
pairing (i.e., neutron superfluidity and/or proton superconductivity, which 
strongly suppress the neutrino emission) in the
densest regions of the neutron star core may be a way to keep the surface 
temperature high enough to be compatible with observations.
Dissipation of the pulsar rotational energy into heat
is another way to control the fast cooling, but detailed modeling of this
effect (Shibazaki \& Lamb 1988; Umeda et al 1993; Van Riper et al 1995)
showed that it is by itself not sufficient.
If the reported temperatures are close to the actual ones and if the
actual temperatures are really due to the cooling of the neutron star
then pairing at very high density seemed to be a mandatory ingredient
for the viability of the fast cooling scenarios.

Recently, however, Potekhin, Chabrier \& Yakovlev (1996) made an important point
by noting that the presence of light elements (H and He) in the upper layers of
the neutron star significantly affects the heat transport in these layers and
results in a higher surface temperature when compared to the case of an iron
envelope.
They also showed that a layer of about $10^{-12}$ \Msol of accreted matter
(possibly deposited during the post-supernova accretion phase)
is sufficient to lead to quite different predictions for cooling models.
I will show here that this hypothesis, when applied to a cooling model with fast
neutrino emission from a kaon condensate (i.e., not as fast as the direct Urca
process, Thorsson et al 1995), and including a reasonable amount of internal
heating, leads to temperatures much higher than previously calculated and that
{\em `agreement with the data' can be obtained without the necessity of baryon
pairing in the inner core where the fast neutrino emission is taking place}, a
result in total opposition to the previous models mentioned above.

I summarize in \S~\ref{sec:data} the observational data on cooling neutron
stars.  My model is described in \S~\ref{sec:input} and the results are
presented in \S~\ref{sec:results}. The last section, \ref{sec:disc}, contains
the discussion and conclusions.

\section{OBSERVATIONAL DATA
         \label{sec:data}}

I summarize in this section the information presently available about
estimates of and upper limits on the effective temperature of cooling neutron
stars and I list in table~\ref{tab:data} all relevant data.
There are several more objects which I omit because the deduced upper limits
on $T_e$ are still too high to be interesting.

The output of cooling models is a luminosity $L^\infty$, i.e., an effective
temperature $T^\infty_e$ related to $L^\infty$ by 
$L^\infty = 4 \pi \sigma R^{\infty \; 2} \; T_e^{\infty \; 4}$
($\sigma$ being the Stefan-Boltzmann constant and $R^\infty$ the neutron star's
radius `at infinity') and the observational problem is thus to deduce an
accurate estimate of $T_e^\infty$ from the data.
The quantities `at infinity' are related to the local ones through the
redshift factor $e^{\phi}$ (= 0.78 for my 1.4 \Msun model) by:
$L^\infty = e^{2 \phi} L$,
$T^\infty = e^{  \phi} T$, and
$R^\infty = e^{- \phi} R$.
In the case of pulsars of goodness 1 to 3 (see table~\ref{tab:data}) the rough
upper limits are based on a comparison with blackbody (BB) spectra.
For the other five objects where there is some evidence in favor of the thermal
origin of the detected photons (see, e.g., \"Ogelman 1995) the question of
the spectrum used to obtain the temperature estimate is crucial.
Model atmospheres (e.g., Romani 1987; Shibanov et al 1992) show that the BB
spectrum may be a very poor approximation;
in particular, magnetized hydrogen atmosphere spectra at high temperatures,
when the atmosphere is wholly ionized, are much harder than BB spectra.
At low temperature ($< 10^6$ K), however, bound-bound and bound-free
absorption may soften substantially the magnetized hydrogen spectrum
and push it toward or even below the BB spectrum (Pavlov \& Potekhin
1995), but no accurate calculations of magnetized spectra have yet
been published in this regime.
In the case of the Vela pulsar, the BB fits give a high temperature,
$T^\infty_{BB} \simeq 1.6 \cdot 10^6$ K, but require a neutron star radius
between 3 -- 4 km (\"Ogelman et al 1993) while fits with magnetized hydrogen
atmosphere spectra, which are expected to be quite accurate at such high
temperature, give a temperature $T^\infty_H \simeq 0.8 \cdot 10^6$ K,
corresponding to a surface temperature $T_H \simeq 10^6$ K, for a neutron
star radius of 10 km, $R^{\infty} = 13.6$ km (Page et al 1996).
However, if the luminosity deduced from the BB fits (with $R^\infty \sim$ 3 -- 4
km) is converted into an effective temperature $T^\infty_e$ for a neutron star
with $R^\infty = 13.6$ km, then $T^\infty_e \sim T^\infty_H$:  
this is the value I report in table~\ref{tab:data}.
The neutron star 0002+6246 (Hailey \& Craig 1995) is very similar to Vela and
I calculate, in the same way, $T^\infty_e$ from $T^\infty_{BB}$.
In the other extreme, the cold Geminga, presently available 
magnetized atmosphere spectra are too
hard, and give very low temperatures which require an enormously large star or,
alternatively, a very small distance (Meyer et al 1993), while BB fits 
(Halpern \& Ruderman 1993) give
higher temperatures which are compatible with emission from the whole surface of
the neutron star at the known distance of 157 (+59, -34) pc (Caraveo et al
1996): I thus choose to use $T^\infty_{BB}$ in this case, i.e., I hope
that $T^\infty_{BB} \sim T^\infty_e$.
The two intermediate cases of PSR 0656+14 and PSR 1055-52 are more delicate: it
is difficult to decide which of the BB or atmosphere spectra give temperature
closer to the actual one.
For PSR 0656+14, I report both values, $T^\infty_{BB}$ (Finley et al 1992)
and $T^\infty_H$ (Anderson et al 1993), the
actual value of $T_e^\infty$ being probably somewhere in-between these, but for
PSR 1055-52 only BB fits have been published (\"Ogelman \& Finley 1993)
so I use $T^\infty_{BB}$ for $T_e^\infty$.

\section{PHYSICS OF THE MODEL
         \label{sec:input}}

I will consider the fast neutrino cooling scenario induced by a kaon condensate
(Brown et al 1988; Page \& Baron 1990).
The dense matter equation of state (EOS) is taken from Friedman \& Pandharipande
(1981), FP hereafter, and I assume that at densities above 
\be 
\rho_{cr}^{K} = 1.1 \times 10^{15} \; \; \rm gm \; cm^{-3} \simeq 4 \cdot \rho_0
\ee
the kaon condensate develops and induces a fast neutrino emission 
\be 
\epsilon_{\nu}^{K} = 10^{24} \; (\rho/\rho_0)^{2/3} \; T_9^6 
             \; \; \rm erg \; s^{-1} \; cm^{-3}
\ee
(Thorsson et al 1995), where $\rho$ is the matter density, $\rho_0 \equiv 2.8
\cdot 10^{14} \; \rm gm \; cm^{-3}$ the nuclear matter density,
and $T_9$ the temperature in units of $10^9$ K.
The critical neutron star mass for the appearance of a kaon condensate is
then 1.24 \Msol.
The modified Urca process and the similar bremstrahlung processes are active at
all densities within the core but, at a temperature of $10^8$ K, the kaon
induced processes are more efficient by five orders of magnitude.
Of upmost importance for the cooling is the presence and extend of baryon
pairing in the core:
the corresponding critical temperatures $T_c$ that I will use are plotted in
figure~\ref{fig:core}.
Notice that the assumed $T_c$'s vanish within the condensate region.
In regions where pairing is present, the specific heat of the paired component
is strongly reduced as well as all neutrino emission processes to which this
component participates.

Surface thermal emission is determined by the effective temperature $T_e$
which is related to the interior temperature $T_b$ through envelope
models.
$T_b$ is traditionally taken at a density $\rho_b = 10^{10} \; \rm gm \; 
cm^{-3}$.
The chemical composition of the envelope can strongly affect the
$T_b - T_e$ relationship:
compared to models with pure iron composition,
models with accreted matter give a higher $T_e$ for a given
$T_b$.
I will use the recent calculations of Potekhin et al (1996) for both pure iron
envelopes and envelopes with a layer of accreted matter, mostly H and He,
of mass $\Delta M$.
For a 1.4 \Msol star with the FP EOS, the total mass of the envelope, i.e., at
densities below $\rho_b$, is about $3 \cdot 10^{-7}$ \Msol.
I neglect here the effects of the magnetic field which are small (Page \& 
Sarmiento 1996).

Besides the neutrino and photon emissions which are energy sinks, rotational
energy can be converted into heat by friction due to differential rotation
of the inner crust neutron superfluid (Alpar et al 1984).
I handle this effect in a simple way by writing the heating rate as
\be
H(t) = J_{44} \cdot 10^{40} \cdot 
       \left( \frac{t+\tau_0}{100 \; {\rm yrs}} \right)^{-3/2} \; \;
       \rm erg \; s^{-1}
\label{equ:heating}
\ee
where $t$ is the pulsar age, $\tau_0 = 300$ yrs a typical spin-down time scale,
and $J_{44}$ the differential angular momentum of the frictionally coupled
crustal neutron superfluid in units of 10$^{44}$ g cm$^2$ rad s$^{-1}$.
This expression assumes a standard spin-down rate from magnetic dipole radiation
and is similar to the one used by Umeda et al (1993) and Shibazaki \& Lamb 
(1988).
This heating is distributed within the superfluid layers of the inner crust.
I adopt the value $J_{44} = 3.1$ which corresponds to moderately strong heating,
compatible with the size of the crust for the FP EOS.

My treatment of other physical ingredients is described in previous works
(Page \& Baron 1990; Page \& Applegate 1992; Page 1994) and
all calculations were performed with a wholly general relativistic Henyey-type
evolutionary code.

\section{RESULTS \label{sec:results}}

Figure~\ref{fig:cool} shows results for the cooling of a 1.4 \Msol neutron star
without and with internal heating and various amounts of accreted mass $\Delta
M$, as well as for a 1.2 \Msol star with heating and $\Delta M = 3 \cdot 
10^{-8}$ 
\Msol.
As seen in figure~\ref{fig:core}, the 1.4 \Msol star contains a kaon condensate,
of mass 0.3 \Msol, while the 1.2 \Msol star has a central density below
$\rho_{cr}$ and thus follows the `standard' cooling scenario.
During the neutrino cooling era, i.e., at age up to about 10$^4$ -- 10$^5$ yrs
when the neutrino luminosity is much higher than the surface photon luminosity,
the nature of the envelope material has no effect on the cooling rate but does
affects the effective temperature: 
models with accreted envelope are warmer than models with iron envelope even
though they all have the same interior temperature (considering separately
models without heating and models with heating, the latter having of course a
higher interior temperature than the formers).
Even such a small amount of accreted mass as $10^{-12}$ \Msol has a very
noticeable effect (Potekhin et al 1996).
When photon cooling takes over, the cooling is accelerated in the cases of
accreted envelopes because of the increased surface emission.
Finally, notice that at the latest ages the cooling is entirely controlled by
the heating, independently of the previous history or envelope composition:
the photon luminosity is simply equal to the heating rate.

Comparing these cooling curves, when both heating and an accreted
envelope are taken into account, with the plotted surface
temperatures, one sees that the `standard' cooling scenario, i.e., M $<$
1.24 \Msol in my model, seems to be appropriate for PSR 1055-52 and
maybe also PSR 0656+14 if the higher temperature estimate is
considered for the latter.
The temperature `measurements' of Vela, NS 0002+6246, Geminga, and PSR
0656+14 if the lower $T_e$ value of the latter is considered, are
below the standard cooling curve but can be fitted very well with the
kaon condensate scenario in presence of heating and with an accreted
envelope of mass larger than $\sim  3 \cdot 10^{-12}$ \Msol.

\section{DISCUSSION AND CONCLUSIONS \label{sec:disc}}

The results of the previous section show that agreement between cooling models
with fast neutrino emission and the best presently available data does not
necessarily require the presence of baryon pairing in the inner core of the 
neutron star: 
the superfluidity $T_c$'s assumed in the calculations (figure~\ref{fig:core})
vanish within the kaon condensate region.
The three basic ingredients needed to obtain temperatures `in agreement with the
data' are: 
1) fast neutrino emission as supplied by a kaon condensate,
2) moderately strong internal heating and 3) the presence of light elements in
the envelope.

There is no consensus on which, if any at all, fast neutrino emission mechanism
is possible in neutron star cores but the kaon condensate is the least 
efficient of all:
a pion condensate or the direct Urca process (Thorsson et al 1995) would shift
downward the cooling curves during the neutrino cooling era and require much
higher heating rates.

The internal heating by friction of the inner crust differentially rotating
neutron superfluid depends on poorly known microscopic parameters (see, e.g.,
Van Riper et al 1995).
A much stronger heating rate than assumed here is theoretically possible but
would conflict with the temperature upper limits of PSR 1929+10 and 0950+08
(see figure~\ref{fig:cool} and Van Riper et al 1995).
Efficient heating, moreover, requires a thick enough crust, i.e., a rather stiff
EOS, and there are arguments for such an EOS precisely in the case of the Vela
pulsar from the analysis of its glitches (Link et al 1992) and from the effect
of gravitational lensing on the amplitude of the pulses seen in its surface
thermal emission (Page \& Sarmiento 1996).
An extreme softening of the EOS in presence of kaon condensation was the result
of the first models (Thorsson et al 1994) but the latest results of Fujii et al
(1996) show that relativistic effects strongly reduce the softening of the EOS
by the condensate: 
a stiff EOS even in the presence of a kaon condensate is thus not unlikely.
Finally, in presence of heating the independence of the late cooling on the
previous history and envelope structure could allow a determination of the
microscopic parameters involved in the problem when detection of thermal
emission from old pulsars will be achieved (Van Riper et al 1995).
Actually, recent Hubble observations (Pavlov et al 1996a) may have detected
the optical -- UV Rayleigh-Jeans tail of the surface thermal emission of
PSR 1929+10 and 0950+08 and indicate surface temperatures of the order of
1 -- 3 $\cdot 10^5$ K for the former and 6 -- 8 $\cdot 10^4$ K for the
latter.
This estimated low temperature of PSR 0950+08 is much below our simple
model prediction but well above any cooling curve which does not include
any heating mechanism:
if confirmed, it would show that {\em heating is required} but that the
simplistic formula \ref{equ:heating} is obviously much too naive.

Due to the enormous uncertainty about the value of the pairing critical
temperature $T_c$ at high density (see, e.g., Page 1994 for a discussion) it is
a relief that the fast cooling scenarios, preferentially the kaon condensate
one, do not necessarily require the presence of superfluidity.
Takatsuka \& Tamagaki (1995) recently emphasized that in presence of a kaon
condensate both neutron and protons would pair in a $^3$P$_2$ state, because of
the high proton fraction, and nucleon pairing is then unlikely unless the 
neutron, or proton, effective mass is unrealistically high.
The relativistic effects described by Fujii et al (1996) imply precisely a
strong reduction of the nucleon effective mass within the condensate region
and disfavor pairing.
However these effects also reduce significantly the softening of the EOS,
implying thicker crust and hence higher heating rate, and the low nucleon
effective masses also reduce $\epsilon_{\nu}^K$ (Brown et al 1988).

Finally, the main question arisen by the results presented above is the
possible presence of light elements at the surface of a neutron star.
The late post-supernova accretion could deposit the needed $10^{-12}$ 
\Msol of light elements.
The only indication I know of about the presence of hydrogen
is in the case of the Vela pulsar when comparing spectral fits with
BB spectra vs. magnetized hydrogen atmosphere spectra as discussed
in \S~\ref{sec:data} which clearly show the inadequacy of blackbody-like
spectra.
Magnetized iron atmosphere spectra (Rajagopal et al 1996) are
blackbody-like when used to fit ROSAT's PSPC low energy resolution
spectra: 
they would imply, for Vela, temperatures and stellar radii similar to
the BB fits, i.e., radii at infinity $R^\infty$ of the order of
3 -- 4 km compared to 14 km for magnetized hydrogen fits.  
Multiwevelength studies (Pavlov et al 1996b) and the future X-ray missions
like SRG, AXAF, and XMM, will help resolve this issue.

I have thus replaced the `magic word' {\em superfluidity} by {\em heating and 
accreted envelope}: 
apparently one of them is needed to make the fast cooling scenarios viable.
The presence of heating and/or of accreted matter will be confronted
by observations in the near future while superfluidity has probably little
chance of getting out of the theoretical realm.

As a last comment I must emphasize that the `classification' of the observed
neutron stars as following the slow (`standard') cooling scenario or
some fast cooling scenario is model dependent.
For example, the present results imply that Geminga and PSR 0656+14
(if the lower $T_e$ is used) belong to the fast cooling scenario but
different assumptions about the extend of baryon pairing in the core
can accommodate them within the `standard' cooling
scenario (Page 1994).
Moreover, Vela and PSR 0002+6246 could also be accommodated within the
`standard' cooling scenario with a radically different assumption about
the efficiency of the modified Urca rate (Schaab et al 1996).

\acknowledgments 

\begin{small}
I thank A. Y. Potekhin for discussions as well as A. Y. P., G. Chabrier
and D. G. Yakovlev for sending me their envelope models before publication.
The idea of this work originated during the ECT* workshop `Physics of
Supernovae and Neutron Stars' (Trento, Italy, 3 -- 14 June 1996).
This work was supported by a UNAM-DGAPA grant No. IN105495 and
a Conacyt grant No. 2127P-E9507.
\end{small}


\begin{figure} 
\caption{\label{fig:core}
         Critical temperatures $T_c$ for neutron and proton pairing in
         the $^1$S$_0$ and $^3$P$_2$ channels used in this work (Page 1994).
         Indicated are also the central densities of
         the 1.2 \Msol and 1.4 \Msol neutron star models used and the crust-core
         and kaon condensate boundaries.}

\caption{\label{fig:cool}
         Thin dotted line:
         Cooling curves for a 1.2 \Msol neutron star (`standard' cooling)
         with heating and accreted envelope ($\Delta M = 3 \cdot 10^{-8}$).
         Thick lines: cooling curves for a 1.4 \Msol neutron star with a kaon
         condensate, various accreted masses and with or without heating.
         The data are from table 1.}
\end{figure}

\clearpage   

\begin{deluxetable}{cccccc}
\tablecaption{Observational data on neutron star surface temperature \tnm{a}
              \label{tab:data}}
\tablehead{
  NS       & Log Age  & Distance &   Log $T_e^\infty$
                                                 & Goodness \tnm{b} 
                                                            & Reference \\
           &  (yrs)   &  (kpc)   &     (K)       &          
&                                }
\startdata
 0531+21   &  3.10    &   2.     &   $<$ 6.25    &    3     
& Becker \& Ashenbach 1995       \nl
 0833-45   &  4.05    &   0.5    &  5.87 -- 5.91 &    5     
& Page et al 1996               \nl
   ''      &   ''     &    ''    &  5.81 -- 5.94 &    5     
& {\"O}gelman et al 1993 \tnm{c} \nl
 0002+6246 & $\sim$ 4 & $\sim$ 3 &  5.72 -- 5.85 &    4     
& Hailey \& Craig 1995 \tnm{c}   \nl
 1706-44   &  4.24    &   1.8    &   $<$ 6.1     &  2 - 3   
& Becker et al 1995             \nl
 2334+61   &  4.61    &   2.5    &   $<$ 6.1     &    2     
& Slane \& Lloyd 1995            \nl
 1916+14   &  4.95    &   1.5    &   $<$ 6.0     &    1     
& Slane \& Lloyd 1995            \nl
 0656+14   &  5.04    &   0.5    &  5.93 -- 5.97 &    5     
& Finley et al 1992              \nl
   ''      &   ''     &   0.3    &  5.70 -- 5.76 &   ''     
& Anderson et al 1993            \nl
 0740-28   &  5.20    &   1.9    &   $<$ 6.0     &    1     
& Slane \& Lloyd 1995            \nl
 1822-09   &  5.37    &   1.0    &   $<$ 5.85    &    1     
& Slane \& Lloyd 1995            \nl
 0630+178  &  5.48    &   0.16   &  5.60 -- 5.70 &    5     
& Halpern \& Ruderman 1993       \nl
 1055-52   &  5.73    &   0.5    &  5.74 -- 5.82 &    5     
& {\"O}gelman \& Finley 1993     \nl
 1929+10   &  6.49    &   0.17   &   $<$ 5.5     &    3     
& Yancopoulos et al 1993        \nl
 0950+08   &  7.23    &   0.13   &   $<$ 5.1     &    2     
& Manning \& Wilmore 1994        \nl
\enddata

\tnt{a}{Listed are: 
        neutron star name; 
        log of age;
        distance; 
        log of effective temperature;
        `goodness' \tnm{b}  of the data;
        reference.}
\tnt{b}{Data `goodness': 
        1: neutron star not detected,
        2: neutron star detected at low count rate which precludes any serious
           analysis of the origin of the photons,
        3: neutron star clearly detected but with evidence that the photons 
           come mostly, or even exclusively, {\em not} from surface thermal 
           emission,
        4: neutron star clearly detected with some spectral evidence about the
           thermal origin of the photons and, finally,
        5: neutron star clearly detected with good spectral evidence about the
           thermal origin of the photons.}
\tnt{c}{The $T_e^\infty$ reported here is really an {\em effective temperature}
        obtained from the luminosity as discussed in 
        section~\protect\ref{sec:data} and not the blackbody temperature 
        $T_{BB}^\infty$ reported by the authors.}

\end{deluxetable}

\clearpage
 
\end {document}
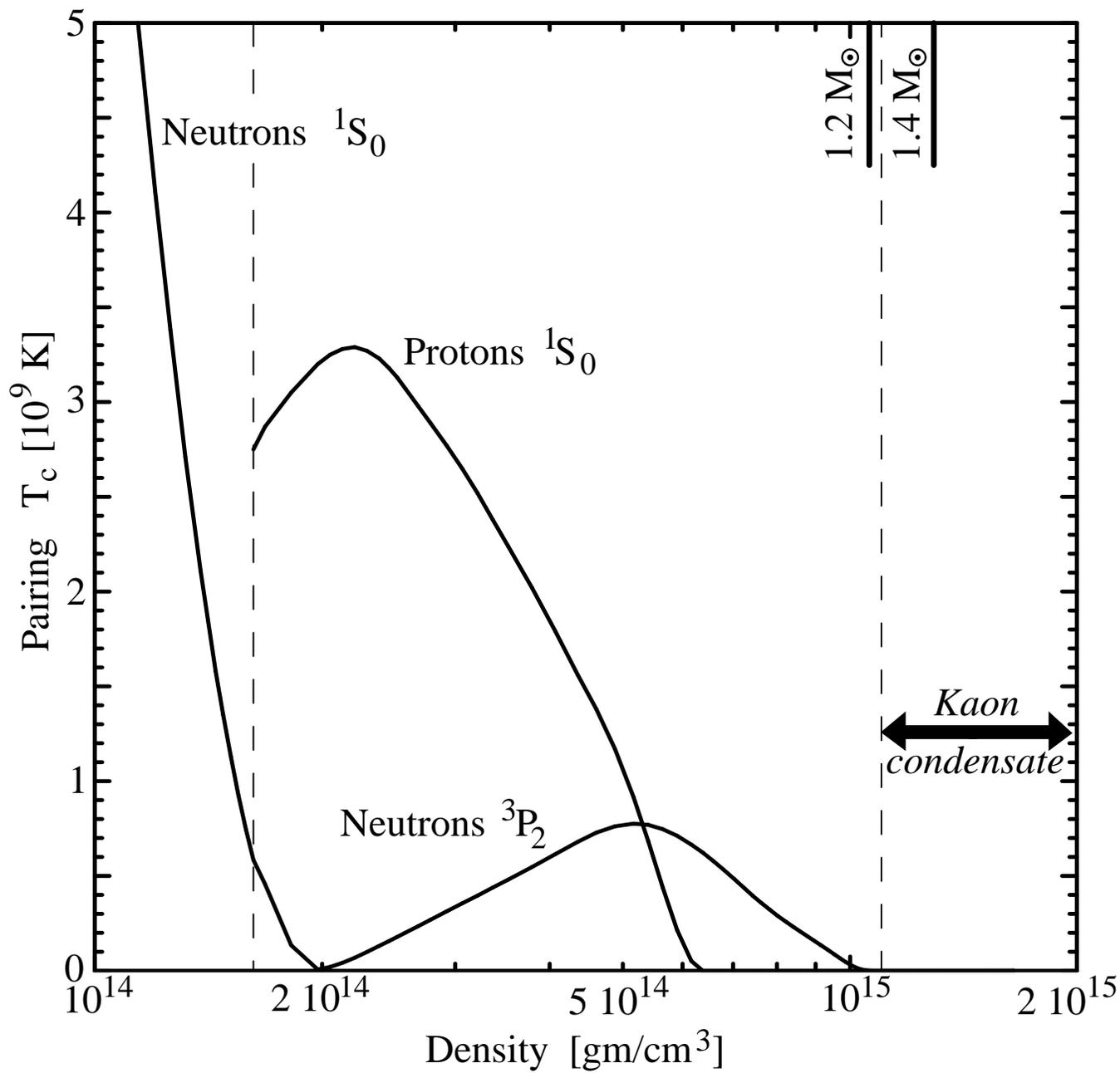
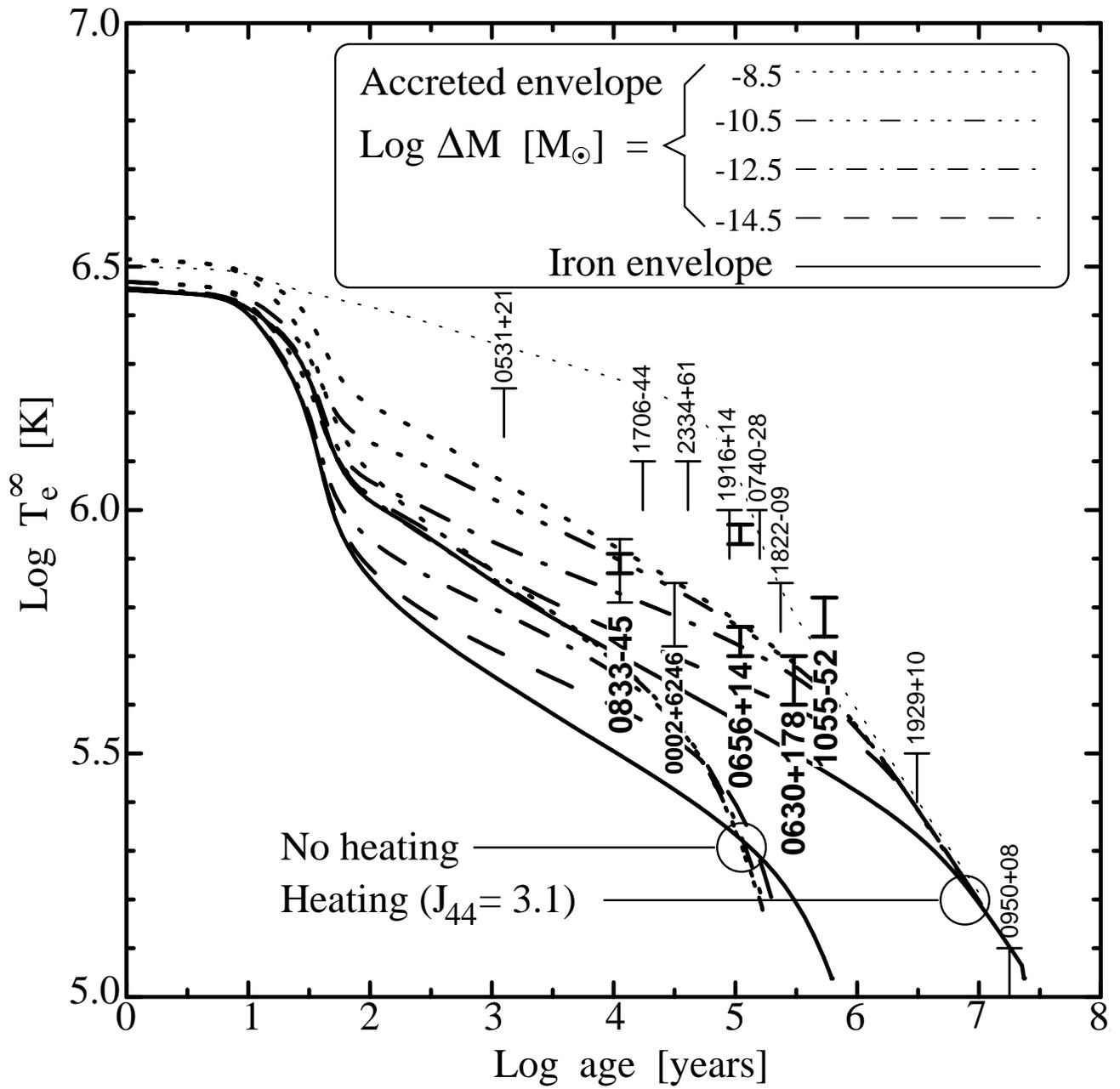
